 \documentclass[final,5p,times,twocolumn]{elsarticle}

\usepackage{subfigure}
\usepackage{amssymb}
\usepackage{lipsum}
\usepackage{amsthm}
\usepackage{graphicx}
\usepackage{amsmath}
\usepackage[colorlinks=true, linkcolor=blue, citecolor=blue, urlcolor=blue]{hyperref}

\journal{Physics Letters B}

\begin{document}

\begin{frontmatter}



    \title{Diving into a holographic multi-band 
 superconductor}


\author[first]{Xing-Kun Zhang}
\ead{zhangxk@nuaa.edu.cn}
\author[first]{Xin Zhao}
\ead{zhaox923@nuaa.edu.cn}
\author[second]{Zhang-Yu Nie}
\ead{niezy@kust.edu.cn}
\author[first,third]{Ya-Peng Hu}
\ead{huyp@nuaa.edu.cn}
\author[first,third]{Yu-Sen An\fnref{label1}}
\fntext[label1]{Corresponding author}
\ead{anyusen@nuaa.edu.cn}
\affiliation[first]{organization={College of Physics, Nanjing University of Aeronautics and Astronautics, Nanjing, 210016, China}}
\affiliation[second]{organization={Center for Gravitation and Astrophysics, Kunming University of Science and Technology, Kunming 650500, China}}
\affiliation[third]{organization={MIIT Key Laboratory of Aerospace Information Materials and Physics,  Nanjing University of Aeronautics and Astronautics, Nanjing, 210016, China}}

\begin{abstract}
In this work, we investigate the interior structure of a holographic multi-band superconductor model. We focus on the holographic superconductor system with two scalar fields which correspond to two s-wave order parameters in the dual condensed matter system. 
We discover a new kind of transition rule among Kasner universe near the black hole singularity which is distinct from the holographic single-band superconductor model. This transition rule is the first generalization of Kasner transition behavior to scenarios involving multiple free parameters, which is useful for uncovering the most general interior structures of hairy black holes. Moreover, we find that the Kasner exponents are sensitive to the details of order parameters in boundary system. These different near singularity structures we found show that the black hole interior plays crucial role in characterizing the boundary dual condensed matter systems.
\end{abstract}




\end{frontmatter}




\section{Introduction}

AdS/CFT correspondence provides a useful tool to understand the strongly coupled condensed matter systems using dual gravitational framework \cite{Maldacena:1997re,Aharony:1999ti}. One prominent example is the holographic superconductor first raised in Ref.\cite{Hartnoll:2008vx,Hartnoll:2008kx} which is an s-wave superconductor model. The physical meaning of this holographic superconductor is as follows. From the bulk side, when gauge field and charged scalar fields are added, the original background without scalar hair (Reissner-Nordstrom black hole) will become unstable to form scalar hairs as we tune the temperature or chemical potential of the gravitational system. This spontaneous generated scalar hair can be understood as the order parameter of superconducting phase transition at the boundary by using AdS/CFT dictionary. Based on Ginzburg-Landau theory, this gravitational model mimics the superconducting phase transition. After the first discovery of this holographic s-wave superconductor model, there are many further generalizations to superconductor models with different types of order parameters, such as p-wave superconductor model \cite{Gubser:2008wv,Cai:2013aca} and d-wave superconductor model \cite{Chen:2010mk,Benini:2010pr}. 

However, instead of one order parameter, investigations on many realistic superconductor systems (such as $MgB_{2}$ and $Sr_{2}Ru O_{4}$ etc) should include 
multiple order parameters  \cite{PhysRevLett.3.552,10.1143/PTP.29.1}. Many theoretical frameworks have also been established in condensed matter theory to understand multi-band superconductor models which have multiple order parameters \cite{PhysRevLett.106.047005,PhysRevB.85.134514}. Besides these theories, AdS/CFT correspondence can also provide an elegant method to describe superconductors with the coexistence of multiple order parameters. Ref.\cite{Basu:2010fa} builds a simple model to describe multi-band superconductor holographically. In this model, there are two charged scalar fields which are coupled to the same $U(1)$ gauge field. It has been shown in Ref.\cite{Basu:2010fa} that there are cases where these two s-wave order parameters coexist. As this work was done in the probe limit,  Ref.\cite{Cai:2013wma} generalized this result to fully back-reacted case and found that the coexistence region will enlarge when increasing the back-reaction. This multi-band model is $\mathrm{s+s}$ type, there are also further generalizations to the coexistence of $\mathrm{s+p}$ order parameters \cite{Nie:2013sda,Nie:2014qma}, $\mathrm{p+p}$ order parameters \cite{Donos:2013woa} and $\mathrm{s+d}$ order parameters \cite{Li:2014wca}. Ref.\cite{Cai:2015cya} gives a comprehensive review of these holographic multi-band superconductor models.

Above references regarding these holographic superconductor models only focus on the black hole exterior region. This is because for the holographic study of condensed matter systems at finite temperature, the boundary state is a mixed state whose gravitational dual is the black hole exterior part. Thus it seems that black hole interior region is not related to these studies. However, the boundary mixed state can be purified to be a pure state. The most well-known purification is canonical purification and the resulting highly entangled pure state is the thermo-field double state which reads 
\begin{equation}
    |TFD\rangle=\frac{1}{\sqrt{Z}}\sum_{i}e^{-\beta E_{i}/2} |E_{i}\rangle_{L}|E_{i}\rangle_{R}.
\end{equation}
where $Z=\sum_{i} e^{-\beta E_{i}}$. Once upon purification, the dual gravitational system of this pure state is the eternal AdS black hole (AdS wormhole) which contains the black hole interior part \cite{Maldacena:2001kr,Maldacena:2013xja}. For different systems with different Hamiltonians, the energy spectrum $E_{i}$ will be different which leads to different thermo-field double state. Certainly, this will also influence the interior structure of corresponding black holes and thus the features of black hole interior should also be important to illustrate the properties of boundary system. 

Based on this consideration, many holographic investigations of boundary system have also been done from black hole interior point of view \cite{Cai:2020wrp,Hartnoll:2020fhc,Hartnoll:2020rwq,An:2021plu,Liu:2021hap,Cai:2021obq,An:2022lvo,Gao:2023zbd,Cai:2023igv,Cai:2024ltu,Carballo:2024hem,Caceres:2023zhl,Hartnoll:2022rdv,Caceres:2021fuw,Xu:2023fad,Gao:2023rqc,Arean:2024pzo,Caceres:2024edr}. For holographic superconductor model, Ref. \cite{Hartnoll:2020fhc} investigated the interior structure of s-wave holographic superconductor proposed in \cite{Hartnoll:2008kx}. They found that the black hole interior structure changes drastically after the superconducting phase transition. The original Cauchy horizon disappears due to the presence of scalar hair. And moreover, they also found a series of critical phenomena inside the horizon which are called "ER collapse" and "Josephson oscillation", these phenomena are most prominent near the critical temperature $T_{c}$. For the region near the singularity, there are also new behaviors. The metric takes the form of anisotropic Kasner universe and there can also exist transitions among these Kasner universe which is similar to the well-known BKL transition  \cite{Belinsky:1970ew,Oling:2024vmq,Damour:2002et}. More notably, black hole interior structures are more sensitive to the features of boundary theories which is different from the  the black hole exterior. For example, the s-wave holographic superconductor model with exponential-type scalar potential and p-wave holographic superconductor model all bear distinct interior structures \cite{Cai:2021obq,An:2022lvo,Cai:2023igv,Cai:2024ltu}. Therefore, the black hole interior structures are very important in the holographic descriptions of condensed matter systems.

For the above progress, the investigations regarding the interior structure all focus on the single order parameter case. There is still no research on the interior structure of holographic multi-band superconductor model. From condensed matter physics, the properties of multi-band superconductor will be very different from single-band superconductor, thus how to characterize this difference holographically is a very interesting question. As the black hole interior structure is a sensitive probe to the boundary model which depends on the models significantly, investigating interior structures of the holographic multi-band superconductor is well motivated which is the main topic of this paper. For simplicity, we will firstly focus on the holographic multi-band superconductor with two charged scalar fields. We find that by incorporating multiple scalar fields, the Kasner geometry and Kasner transition behavior near the black hole singularity will present new features compared to single scalar field case. These new features are not only important to characterize the properties of multi-band superconductor but also constitute an important part towards finding the most general interior structure of hairy black holes.

The structures of this letter are as follows. In Sec.\ref{review}, we first review the holographic multi-band  superconductor with two competing scalar fields which have different mass and charge. We will show the three different types of coexistence behavior for different choice of charge parameters. In Sec.\ref{interior}, we will investigate the interior structure of this holographic multi-band  superconductor. We investigate both the near horizon region and near singularity region and describe its difference between the holographic single-band superconductor model. In Sec.\ref{cond}, we conclude our paper and give some future directions.


\section{Holographic multi-band superconductor with two scalar fields:} \label{review}
In this section, we will first give a brief introduction to the holographic multi-band superconductor model where two order parameters can coexist and compete. We will only focus on the case where two competing condensates are both scalar. To model the two s-wave orders on the boundary, two scalar fields are introduced in the bulk theory. Thus the action  takes following form
\begin{equation}
        S=\frac{1}{2\kappa^{2}}\int d^{4}x\sqrt{-g}[R+\frac{6}{L^{2}}-\frac{1}{4}F_{\mu\nu}F^{\mu\nu}+\sum_{k=1}^{2}(-|\nabla \psi_{k}-ie_{k}A\psi_{k}|^{2}-m_{k}^{2}|\psi_{k}|^{2})],
\end{equation}
where the two scalar fields are minimally coupled to the same $U(1)$ gauge field and there is no direct coupling between them. $e_{k}$ and $m_{k}$ are the charge and mass of scalar field $\psi_{k}$, $\kappa^{2}=8\pi G $ is the Planck length and  $F_{\mu\nu}=\nabla_{\mu}A_{\nu}-\nabla_{\nu}A_{\mu}$ is the electro-magnetic field strength. 


For time-independent s-wave superconductor model, the static and spherically symmetric metric ansatz should be imposed as
\begin{equation}\label{ansatz}
    ds^{2}=\frac{1}{z^{2}}(-f e^{-\chi}dt^{2}+\frac{dz^{2}}{f}+dx^{2}+dy^{2}),
\end{equation}
moreover, the corresponding matter field ansatz read
\begin{equation}
    \psi_{1}=\psi_{1}(z), \quad \psi_{2}=\psi_{2}(z), \quad A=\phi(z)dt.
\end{equation}
With these special ansatz, by varying the action with respect to the scalar field $\psi_{k}$, electromagnetic field $A_{\mu}$ and the metric field $g_{\mu\nu}$, the equations of motion read 

\begin{equation}\label{eom1}
    z^{2}e^{\chi/2}(z h \psi_{1}')'=\frac{m_{1}^{2}\psi_{1}}{z^{2}}-\frac{e^{\chi/2}e_{1}^{2}\phi^{2}\psi_{1}}{hz^{3}}\ ,
\end{equation}
\begin{equation}\label{eom2}
    z^{2}e^{\chi/2}(z h \psi_{2}')'=\frac{m_{2}^{2}\psi_{2}}{z^{2}}-\frac{e^{\chi/2}e_{2}^{2}\phi^{2}\psi_{2}}{h z^{3}}\ ,
\end{equation}
\begin{equation}\label{eom3}
    z^{2}(e^{\chi/2}\phi')'=\frac{2\phi(e_{1}^{2}\psi_{1}^{2}+e_{2}^{2}\psi_{2}^{2})}{h z^{3}}\ ,
\end{equation}
\begin{equation}\label{eom4}
\chi'=z(\psi_{1}'^{2}+\psi_{2}'^{2})+\frac{\phi^{2}}{h^{2}z^{5}}(e_{1}^{2}\psi_{1}^{2}+e_{2}^{2}\psi_{2}^{2})\ ,
\end{equation} 
\begin{equation}\label{eom5}
4z^{4}e^{\chi/2}h'=-12+2m_{1}^{2}\psi_{1}^{2}+2m_{2}^{2}\psi_{2}^{2}+e^{\chi}z^{4}\phi'^{2}\ ,
\end{equation}
where $h=\frac{fe^{-\chi/2}}{z^{3}}$ is introduced for convenience. The first two equations are scalar field equations and the third equation is Maxwell equation. The last two equations are two independent Einstein equations. Solving these five coupled differential equations with suitable boundary condition will holographically model the multi-band superconductor system. The parameter $e_{1}$,$e_{2}$ controls the back-reaction of the scalar field to the background. It has been found that the strength of back-reaction will strongly influence the coexistence behavior of two scalar order parameters. To be more specific, increasing the back-reaction (decreasing the charge) will enlarge the coexistence region \cite{Cai:2013wma}. \footnote{Note that infinite charge limit corresponds to probe limit where back-reaction effect is switched off.}

For the convenience to perform numerics, mass parameters of two scalar fields are chosen to be $m_{1}^{2}=0$ and $m_{2}^{2}=-2$. For the spacetime having asymptotic AdS boundary, the general falloff of matter fields and metric fields near the boundary takes the form 
\begin{equation}
\begin{aligned}
    & \psi_{1}= \psi_{1}^{(0)} z^{\Delta_{1-}}+\psi_{1}^{(1)} z^{\Delta_{1+}}+...\ , \quad \psi_{2}= \psi_{2}^{(0)} z^{\Delta_{2-}}+\psi_{2}^{(1)} z^{\Delta_{2+}}+...\ ,\\& f = \frac{1}{z^{2}}+...\ , \quad \chi = 0+...\ , \quad \phi = \mu-\rho z+...\ .
    \end{aligned}\label{falloff}
\end{equation}
By plugging these falloffs into the Klein-Gordon equation Eq.(\ref{eom1}) and Eq.(\ref{eom2}), mass-dimension relation in four dimension can be derived as
\begin{equation}
    \Delta_{1}(\Delta_{1}-3)=m_{1}^{2}, \quad \Delta_{2}(\Delta_{2}-3)=m_{2}^{2}.
\end{equation}
It can be easily solved that for the first scalar field, the conformal dimension is $\Delta_{1-}=0$ and $\Delta_{1+}=3$ and for the second scalar field, the conformal dimension is $\Delta_{2-}=1$ and $\Delta_{2+}=2$.  
Thus the expansion of scalar fields near boundary are correspondingly 
\begin{equation}   \psi_{1}=\psi_{1}^{(0)}+\psi_{1}^{(1)}z^{3}+...\ ,\quad \psi_{2}=\psi_{2}^{(0)}z+\psi_{2}^{(1)}z^{2}+... \ .
\end{equation}
Standard quantization procedure is chosen in this paper where $\psi^{(0)}$ is identified as the source and $\psi^{(1)}$ as the expectation value. To build a superconductor model, scalar condensate should be spontaneously generated thus the boundary source terms are set to be zero 
\begin{equation}\label{bdc}
    \psi_{1}^{(0)}=0, \quad \psi_{2}^{(0)}=0.
\end{equation}

To avoid coordinate singularity at horizon, one can expand the fields near the horizon and plug the series into the equations of motion. It can be found that there are five independent parameters at the horizon $\{z_{h},\psi_{1}(z_{h}),\psi_{2}(z_{h}),\phi'(z_{h}),\chi(z_{h})\}$. Note that there are scaling symmetries associated to the equations of motion
\begin{equation}
    e^{-\chi} \to \lambda^{2}e^{-\chi}, \quad \phi \to \lambda \phi, \quad t \to \lambda^{-1}t ;
\end{equation}
\begin{equation}\label{scale2}
    z\to \lambda z, \quad (t,x,y) \to \lambda^{-1}(t,x,y), \quad f \to \lambda^{2}f, \quad \phi \to \lambda \phi.
\end{equation}
When performing numerics, one can first set  $z_{h}=1$ and $\chi(z_{h})=0$ and then use the first scaling symmetry to set $\chi(0)=0$, \footnote{This condition guarantees the asymptotic condition where the boundary time is the same as bulk time.} thus only three independent parameters $\{\psi_{1}(z_{h}),\psi_{2}(z_{h}),\phi'(z_{h})\}$ remain. For this boundary value problem, we should invoke the shooting method to numerically solve the differential equations (\ref{eom1}-\ref{eom5}) with boundary conditions (\ref{bdc}). The specific procedure of numerical computation is as follows: firstly under prescribed $\phi'(z_{h})$, the fourth-order Runge-Kutta method was selected to numerically determine the boundary value of $\psi_{1}^{(0)}$ and $\psi_{2}^{0}$ under specified choice of shooting parameters $\{\psi_{1}(z_{h}),\psi_{2}(z_{h})\}$. If the boundary conditions are not satisfied, we use the Newton-Raphson iterative method to search for $\{\psi_{1}(z_{h}),\psi_{2}(z_{h})\}$ which matches boundary conditions. After getting the parameters on the horizon, the coupled differential equations (\ref{eom1}-\ref{eom5}) are directly solved and the expectation value of order parameter $\psi_{1,2}^{(1)}$ are extracted.

Moreover, for the expansion of electromagnetic field at boundary in Eq.(\ref{falloff}), $\mu$ is the chemical potential and $\rho$ is the charge density. We choose to work in grand canonical ensemble thus the chemical potential should be kept fixed. We will fix the chemical potential $\mu=1$ by using the second scaling symmetry Eq.(\ref{scale2}) in this work. 

\begin{figure}[h]
    \centering  \includegraphics[width=0.45\textwidth]{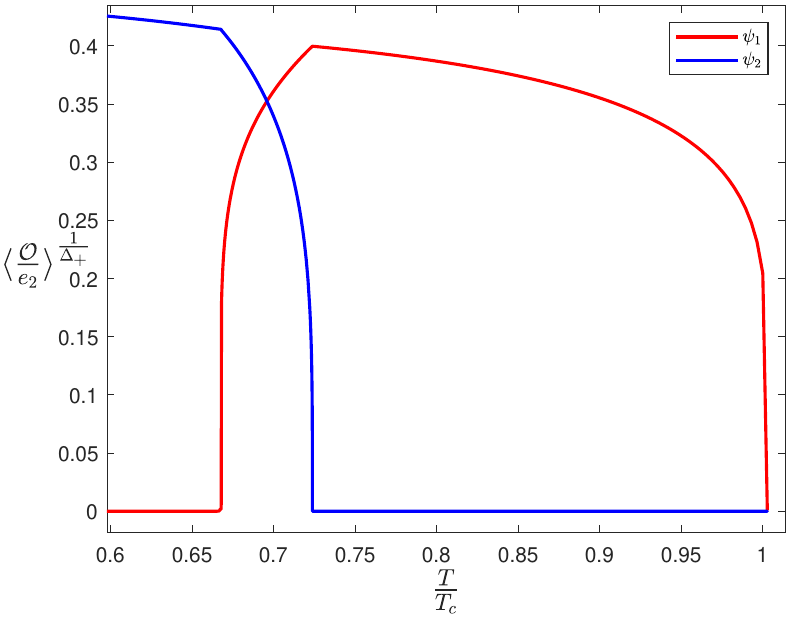}
    \caption{Phase C of multiple order parameters in holographic superconductor. This figure is plotted for parameter $e_{2}=4$ and $e_{1}/e_{2}=1.95$.The red curve corresponds to $\psi_{1}$ while the blue curve corresponds to $\psi_{2}$. We fix chemical potential $\mu=1$ in this figure.It can be seen that when the second order parameter appears, it suppresses the first order parameter.}
    \label{qx}
\end{figure}

\begin{figure*}[h]
    \centering
    \subfigure[]{\includegraphics[width=0.5\textwidth]{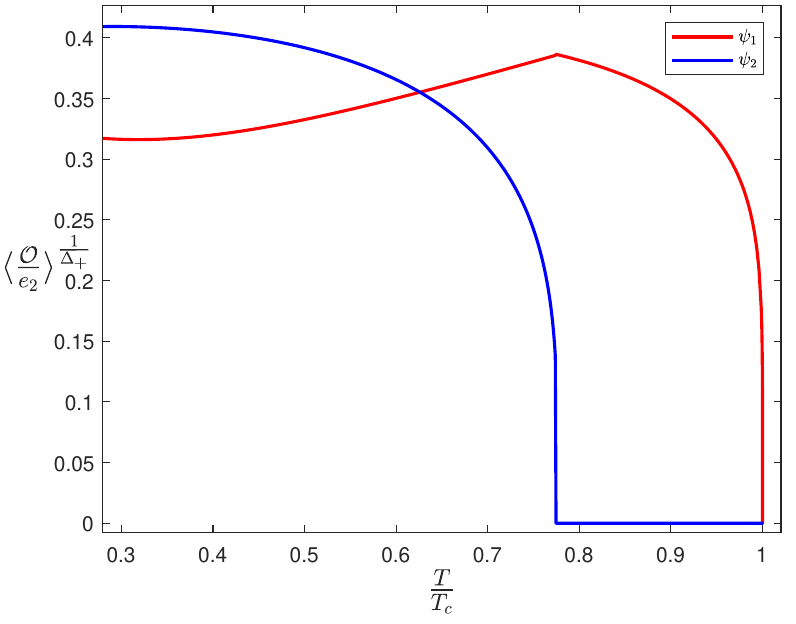}}\hfill
    \subfigure[]{\includegraphics[width=0.48\textwidth]{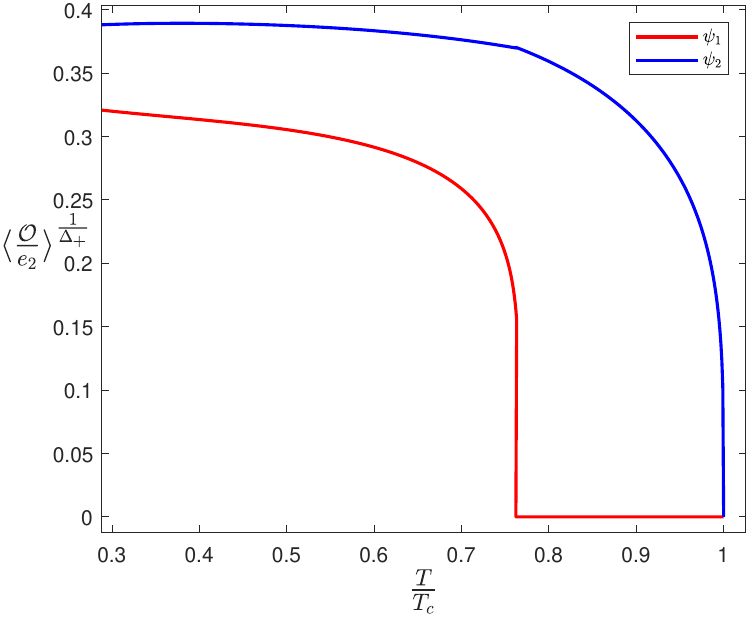}}
    \caption{Multiple order parameters in holographic superconductor for larger back-reaction. Left Panel:Phase A with $e_{2}=2$ and $e_{1}/e_{2}=1.95$. Right Panel:Phase B with $e_{2}=1.5$ and $e_{1}/e_{2}=1.9$. The red curve corresponds to $\psi_{1}$ while the blue curve corresponds to $\psi_{2}$. We fix chemical potential $\mu=1$ in this figure. It can be seen that by increasing the back-reaction, the suppressive effect of second order parameter will be less dramatic and the two order will always coexist.}
    \label{qx1}
\end{figure*}

From the above prescriptions, the condensate behaviors can be deduced. At high temperature, the normal phase is dominant which is RN-AdS black hole where two scalar field vanishes $\psi_{1}=\psi_{2}=0$. After decreasing temperature, the spontaneous symmetry breaking happens and scalar order parameter appears. It is interesting to note that when there are two scalar fields, when further lowering the temperature, second order parameter will appear which have significant influence on the first order parameter. Most notably, it was plotted in Fig.\ref{qx} that second order parameter will suppress the first order parameter \footnote{Note that we rescale the condensate value using $e_{2}$ in this figure to match the notation in Ref.\cite{Cai:2013wma}.} and by further decreasing temperature, the second order parameter will still increase and the first order parameter will decrease to zero. This case is called phase-C in Ref.\cite{Cai:2013wma}.
In this case, the two order parameters have strong competition between each other. 
Furthermore, by increasing the back-reaction (decreasing the parameter $e_{1}$ or $e_{2}$), it was found in Ref.\cite{Cai:2013wma} that strong back-reaction will enlarge the coexistence region which makes two order parameters always coexist. Ref.\cite{Cai:2013wma} found two possible phases, one is shown in the left panel of Fig.\ref{qx1} for the parameter $e_{2}=2$ and $e_{1}/e_{2}=1.95$ which is called phase-A in \cite{Cai:2013wma} , and another is shown in the right panel of Fig.\ref{qx1} for the parameter choice $e_{2}=1.5$ and $e_{1}/e_{2}=1.9$ which is called phase-B in \cite{Cai:2013wma}. 

These condensate behaviors imply that the microscopic physics behind this superconductor model will be much more complex than the single order parameter case. Due to the close connection between black hole interior structure and boundary system, it is natural to expect that the coexistence of two order parameters can also significantly influence the black hole interior. Thus in the next section, we will investigate the corresponding black hole interior structure to show this distinctiveness. 

\section{Interior structure of holographic multi-band superconductor:}\label{interior}
By directly solving the equations of motion with specific shooting parameters, the condensate behaviors and exterior region of holographic superconductor can be known. However, besides the exterior part, it is also interesting to directly go beyond the horizon to solve the interior region of holographic superconductor which presents many interesting features \cite{Hartnoll:2020rwq}.  Therefore, we will investigate the interior structure of holographic multi-band superconductor in this section. For the purpose of simplicity, we only investigate these phenomena in Phase-C case, behaviors in Phase-A and Phase-B case are similar.
\subsection{General structures:}
\begin{figure*}[!h]
    \centering
    \subfigure[Plot of metric function $-h(z)$]{\includegraphics[width=0.52\textwidth]{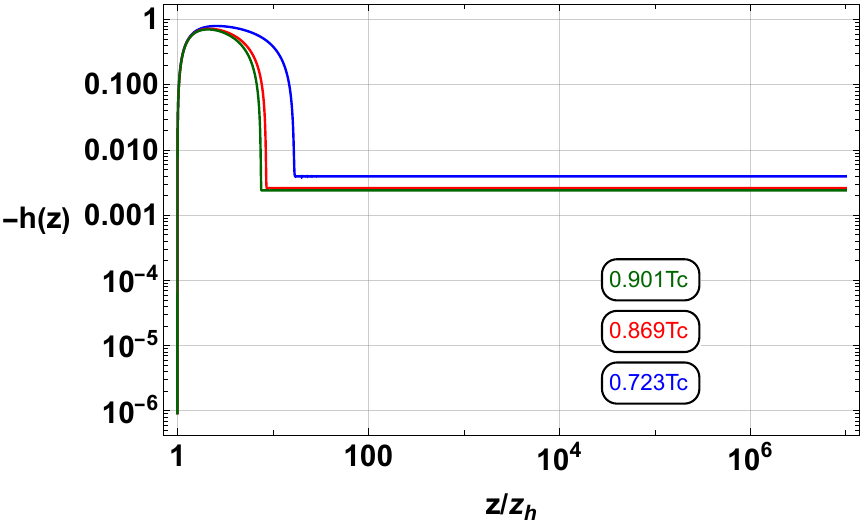}}\hfill
    \subfigure[Plot of gauge field $\phi(z)$]{\includegraphics[width=0.48\textwidth]{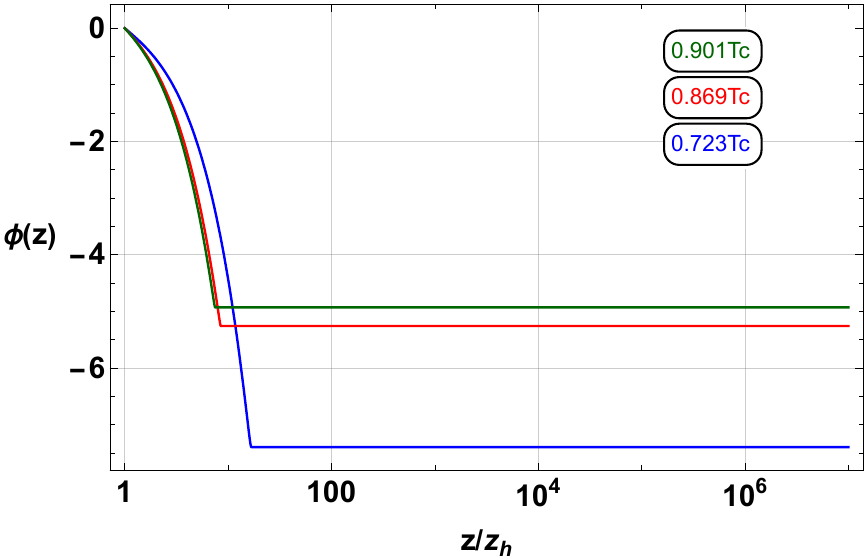}}
    \caption{We see $h(z)$ is an $O(1)$ quantity in the deep interior which means the singularity is space-like. }
    \label{Fig:ex-solutionh}
\end{figure*}

For charged black hole with scalar hair, it can be shown that when the scalar hair forms, no matter how small it is, it can induce strong non-linear effect which removes the Cauchy horizon\cite{Hartnoll:2020fhc,Cai:2020wrp,An:2021plu}. There are many proofs of the no Cauchy horizon theorem by using conserved charge method and energy condition method \cite{Cai:2020wrp,An:2021plu,Yang:2021civ}. The proof is valid for arbitrary numbers of minimally coupled scalar fields, thus there is no Cauchy horizon in our multi-band superconductor model and the black hole singularity becomes space-like when scalar condensate forms. Absence of Cauchy horizon can also be corroborated by the  numerical result, we plot the spacetime interior structure and gauge field behavior inside black hole in Fig.\ref{Fig:ex-solutionh} to show this. It is easy to see that $h(z)=0$ and $\phi(z)=0$ have only one root which means the black hole has only one horizon.

\subsection{Interior structure near the horizon:}
Besides the absence of Cauchy horizon, there are also more novel behaviors inside the corresponding scalar hairy black hole. Notable effects include the ER collapse and Josephson oscillation of scalar field \cite{Hartnoll:2020fhc}. Thus we will first investigate them for multi-band model. 

\begin{figure}[!h]
\centering
\includegraphics[width=0.45\textwidth]{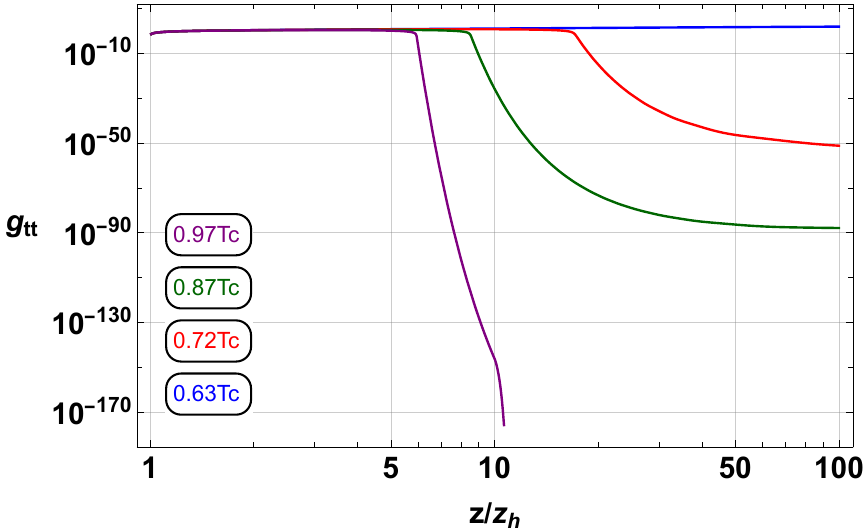}
\caption{ER collapse behavior inside the holographic superconductor, we see that the ER collapse is less dramatic as we lower the temperature away from critical temperature }\label{gtt}
\end{figure}

Near the horizon, similar to the results in Ref.\cite{Hartnoll:2020fhc}, there will still be sudden decrease of metric function $g_{tt}$ which is called ER collapse. We plot the ER collapse behavior in different temperature in Fig.\ref{gtt}. It can be seen that the ER collapse will be less dramatic as we decrease the temperature away from $T_{c}$.This means that the collapse of $g_{tt}$ is a critical phenomenon which is only notable near the critical temperature $T_{c}$.

\begin{figure*}[ht]
    \centering
    \subfigure[Scalar oscillation behavior for $T=0.97T_{c}$]{\includegraphics[width=0.33\textwidth]{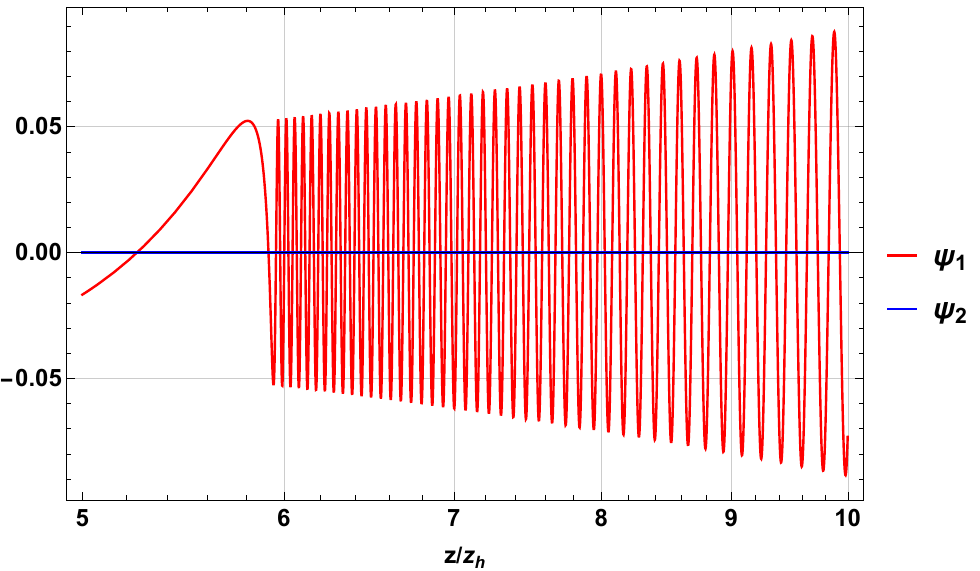}}\hfill
    \subfigure[Scalar oscillation behavior for $T=0.72T_{c}$]{\includegraphics[width=0.33\textwidth]{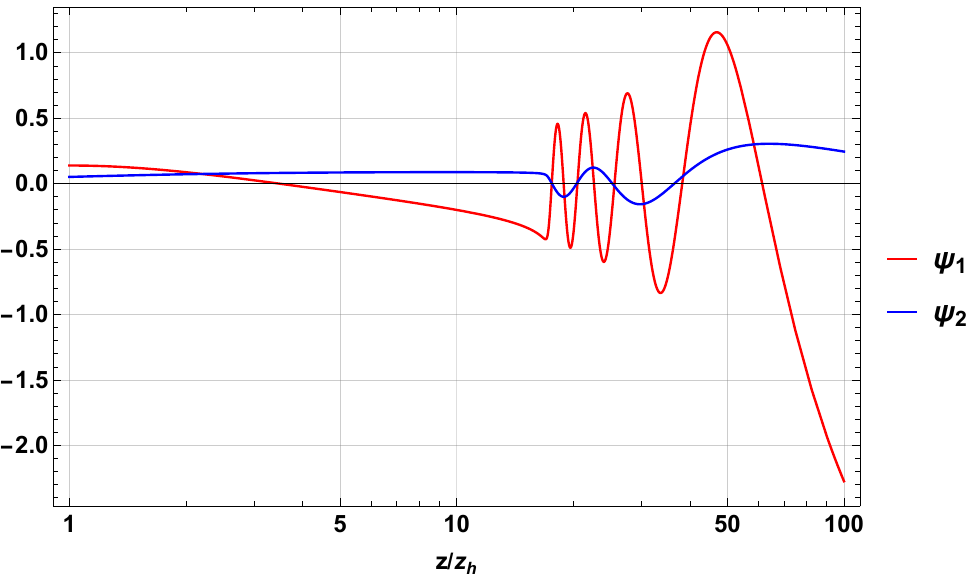}}
    \subfigure[Scalar oscillation behavior for $T=0.68T_{c}$]{\includegraphics[width=0.32\textwidth]{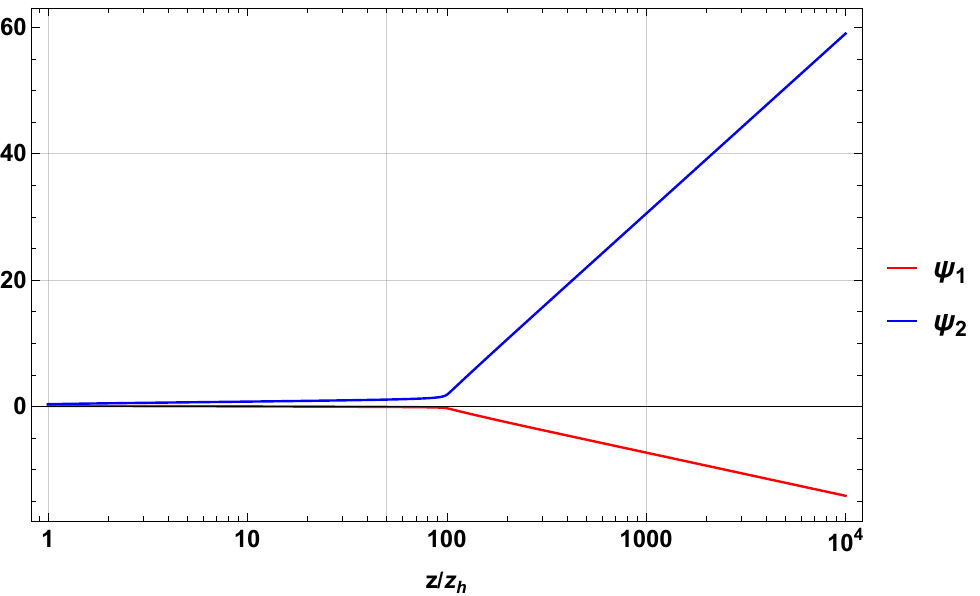}}
    \caption{Josephson oscillation for different temperature. When lowering temperature, the oscillation gradually disappears as ER collapse gradually diminishes.}
    \label{josone}
\end{figure*}

ER collapse is closely related to the phenomenon called Josephson oscillation. The scalar field inside black hole can have oscillations right after ER collapse happens. This can be found in terms of the equation of motion of scalar field \footnote{The two scalar fields have the same form of equation of motion,$\psi$ here denotes either $\psi_{1}$ or $\psi_{2}$.} which reads 
\begin{equation}
    \psi''+\frac{h'}{h}\psi'+\frac{\psi'}{z}=-\frac{q^{2}\phi^{2}\psi}{z^{6}h^{2}}+\frac{m^{2}\psi}{z^{5}h}e^{-\chi/2}.
\end{equation}
After ER collapse, $e^{-\chi/2}$ is vastly small thus the mass term and $h'/h$ term can be neglected. Moreover, it can be checked numerically in Fig.\ref{Fig:ex-solutionh} that the $\phi^{2}/h^{2}$ term is approximately a constant after ER collapse. After using these approximations, the scalar field can be solved as 
\begin{equation}\label{bessel}
    \psi=c_{J}J_{0}(\frac{\phi}{2hz^{2}})+c_{Y}Y_{0}(\frac{\phi}{2hz^{2}}),
\end{equation}
where $c_{J}$,$c_{Y}$ are two integration constants and $J_{0}$,$Y_{0}$ are first kind and second kind Bessel function respectively. Eq.(\ref{bessel}) is a highly oscillating function which is called "Josephson oscillation" in Ref.\cite{Hartnoll:2020fhc} to show its close connection to Josephson effect in superconductor theory. This analytical argument can also be corroborated by the numerical result. We plot the Josephson oscillation behavior of holographic multi-band  superconductor in Fig.\ref{josone}. For the temperature close to the critical temperature, we see that there is one scalar hair which oscillates rapidly. As we lower the temperature, two things happen. Firstly, the ER collapse becomes less dramatic which leads to slower oscillation of scalar field. Secondly, below a certain temperature ($T=0.72360 T_{c}$ in our case), second scalar field appears. As was shown in the middle panel of Fig.\ref{josone}, for temperature $T=0.72T_{c}$, there will be two oscillating scalar fields. Further decreasing the temperature will eliminate the Josephson oscillation which is shown in the right panel of Fig.\ref{josone}. 

From the above results, we conclude that the interior structure of holographic multi-band superconductor near the horizon is quite similar to the results for single-band case. However, we will show in the following that the physical properties near the singularity will be quite different. 

\subsection{Interior structure near the singularity:}
\begin{figure*}[h]
    \centering
    \subfigure[Behavior of $z\psi_{1}'$]{\includegraphics[width=0.5\textwidth]{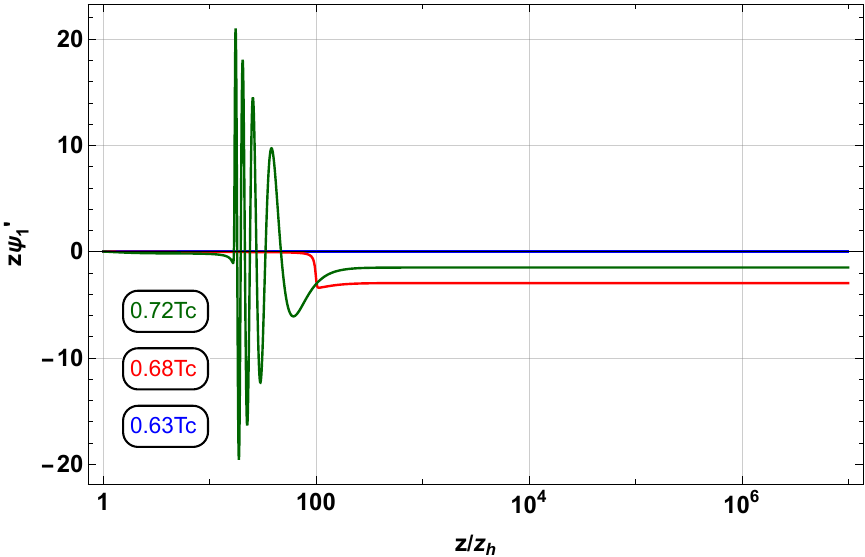}}\hfill
    \subfigure[Behavior of $z\psi_{2}'$]{\includegraphics[width=0.49\textwidth]{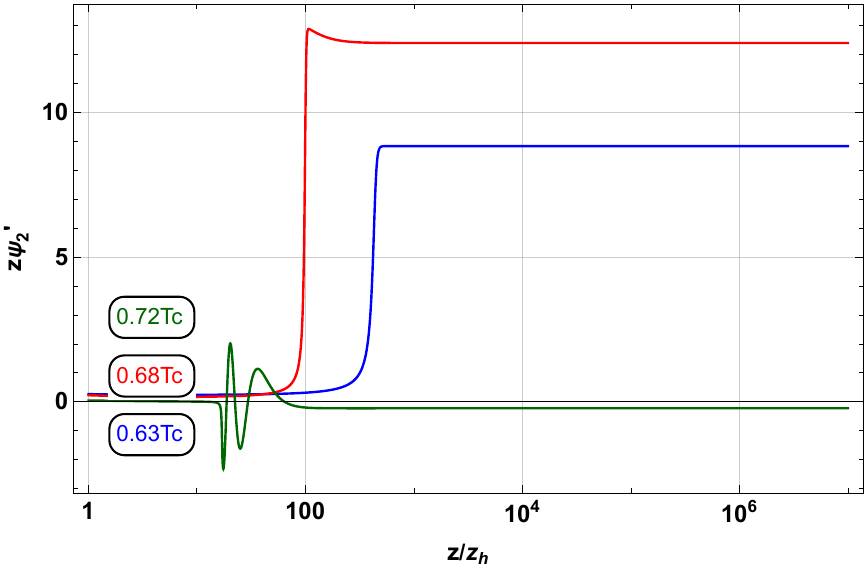}}
    \caption{The scalar field behavior inside black hole. $z\psi'$ approaches constant in the deep interior implies that the metric enters into the Kasner epoch. }
    \label{zpsii}
\end{figure*}

From the behavior of $z\psi'$ in Fig.\ref{zpsii}, we see that  the scalar field will change logarithmically with respect to $z$ (which is equivalent to $z\psi'_{1,2}$ approach constants) in the deep interior. This behavior indicates that the system enters into the new epoch.
We can check numerically that in this region, the mass term and charge term in the equations of motion can be neglected, and the equation of motion for $\psi_{1}$,$\psi_{2}$ and $\chi$ can be vastly simplified as follows 
\begin{equation}\label{dfss}
    \psi_{1}'' \approx -\frac{1}{z}\psi_{1}', \quad \psi_{2}'' \approx -\frac{1}{z}\psi_{2}',\quad \chi'\approx z(\psi_{1}'^{2}+\psi_{2}'^{2}).
\end{equation}
After this approximation, the scalar fields and the function $\chi$ can be solved as 
\begin{equation}
    \psi_{1} \sim \alpha_{1} \ln z, \quad \psi_{2} \sim \alpha_{2} \ln z,\quad \chi \sim (\alpha_{1}^{2}+\alpha_{2}^{2}) \ln z 
    \label{chik}
\end{equation}
where $\alpha_{1}$,$\alpha_{2}$ are two integration constants. 
As can be verified from numerical results in Fig.\ref{Fig:ex-solutionh} posteriorly, $h=\frac{f e^{-\chi/2}}{z^{3}}$ is a $\mathcal{O}(1)$ quantity, thus we find the metric function $f$ will have following form
\begin{equation}
    f \sim z^{3+\frac{1}{2}\alpha_{1}^{2}+\frac{1}{2}\alpha_{2}^{2}}.
\end{equation}
By using the metric ansatz in Eq.(\ref{ansatz})
, the metric component $g_{zz}$ at large $z$ reads 
\begin{equation}
    g_{zz} \sim \frac{1}{z^{5+\frac{1}{2}\alpha_{1}^{2}+\frac{1}{2}\alpha_{2}^{2}}}.
\end{equation}
By transforming to proper time $\tau$ using the relation $d\tau=\sqrt{g_{zz}}dz$,
we found that the metric near the singularity reads 
\begin{equation}
    ds^{2}=-d\tau^{2}+\tau^{2p_{t}}dt^{2}+\tau^{2p_{x}}dx^{2}+\tau^{2p_{y}}dy^{2},
\end{equation}
with Kasner exponent 
\begin{equation} p_{t}=\frac{\alpha_{1}^{2}+\alpha_{2}^{2}-2}{\alpha_{1}^{2}+\alpha_{2}^{2}+6},\quad p_{x}=p_{y}=\frac{4}{\alpha_{1}^{2}+\alpha_{2}^{2}+6}.
\end{equation}
Moreover, we can define $p_{\psi}$ as $\psi \sim -p_{\psi} \ln \tau$ and find that 
\begin{equation}
    p_{\psi_{1}}=\frac{4\alpha_{1}}{\alpha_{1}^{2}+\alpha_{2}^{2}+6}, \quad   p_{\psi_{2}}=\frac{4\alpha_{2}}{\alpha_{1}^{2}+\alpha_{2}^{2}+6}.
\end{equation}
Thus with two independent scalar fields, the Kasner exponents should be determined jointly by the two parameters $\alpha_{1}$ and $\alpha_{2}$. It can be easily checked that the above Kasner exponents satisfy the following two interesting relations
\begin{equation}
    p_{t}+p_{x}+p_{y}=1,\quad p_{t}^{2}+p_{x}^{2}+p_{y}^{2}+p_{\psi_{1}}^{2}+p_{\psi_{2}}^{2}=1.
\end{equation}

In this Kasner regime, the Maxwell equation can also be easily solved, where the electro-magnetic field reads 
\begin{equation}
    \phi \sim \phi_{0}+E_{0} z^{1-\frac{1}{2}\alpha_{1}^{2}-\frac{1}{2}\alpha_{2}^{2}},
\end{equation}
with $\phi_{0}$ and $E_{0}$ two integration constants. It can be seen that for $\alpha_{1}^{2}+\alpha_{2}^{2}<2$, the electro-magnetic field will grow for large $z$. Thus for the case $\alpha_{1}^{2}+\alpha_{2}^{2}<2$, the accumulated electro-magnetic field can become important which makes the Kasner region unstable. The unstable Kasner region will transit to another stable Kasner region. So in the following, we should take this fact into account and investigate the Kasner transition behaviors for multi-band holographic superconductor.

In the Kasner region, as verified posterity that the terms involving $e_{1}$ and $e_{2}$ and $m$ in all the equations of motion can be neglected, the Maxwell equation (\ref{eom3}) and Einstein equation (\ref{eom5}) can be solved as 
\begin{equation}\label{kphi}
  \phi=\phi_{0}+E_{0}\int e^{-\chi/2}dz,\quad  h=\frac{E_{0}^{2}}{4}\int e^{-\chi/2}dz- \frac{1}{c}
\end{equation}
where $\phi_{0}$, $E_{0}$ and $c$ are three integration constants. 
Moreover, by re-writing scalar field as 
\begin{equation} \label{kpsi}
    \psi_{1}=\int \frac{\alpha_{1}(z)}{z}dz,\quad \psi_{2}=\int \frac{\alpha_{2}(z)}{z}dz ,
\end{equation}
the equation of $\chi$ in Eq.(\ref{dfss}) becomes 
\begin{equation} \label{kchi}
    \chi'=\frac{\alpha_{1}(z)^{2}+\alpha_{2}(z)^{2}}{z}.
\end{equation}
Plugging Eq.(\ref{kphi}-\ref{kchi}) into the scalar field equations, the scalar field equations become 
\begin{equation} \label{inv1} \alpha_{2}^{2}+\alpha_{1}^{2}+2z\frac{\alpha_{1}''}{\alpha_{1}'}-4z \frac{\alpha_{1}'}{\alpha_{1}}=0,
\end{equation}
\begin{equation}  \label{inv2} \alpha_{1}^{2}+\alpha_{2}^{2}+2z\frac{\alpha_{2}''}{\alpha_{2}'}-4z \frac{\alpha_{2}'}{\alpha_{2}}=0.
\end{equation}
Note that when one of the scalar fields vanishes, the above coupled differential equations becomes one single differential equation
\begin{equation}
\alpha^{2}+2z\frac{\alpha''}{\alpha'}-4z \frac{\alpha'}{\alpha}=0,
\end{equation}
which can be directed solved as 
\begin{equation}
    (\alpha-\alpha_{0})^{-2/(2-\alpha_{0}^{2})}(\frac{2}{\alpha_{0}}-\alpha)^{-1/(1-2/\alpha_{0}^{2})}\alpha=\frac{z_{in}}{z},
\end{equation}
where $z_{in}$ is an integration constant. This solution tells us that if the first Kasner epoch has the feature $\alpha_{0}^{2}<2$, when increasing $z$, $\alpha(z)$ should transit from $\alpha_{0}$ to $2/\alpha_{0}$. This is the so called "Kasner inversion" behavior which is first introduced in Ref.\cite{Hartnoll:2020fhc}.\footnote{The Kasner transition behaviors are different for different matter fields or gravitational theories, for more discussions see Ref.\cite{An:2022lvo,Cai:2024ltu,Caceres:2024edr,Bueno:2024fzg}. } However, for the multi-band model in the presence of second scalar field, the simple Kasner inversion behavior will change significantly. In the multi-band case, the two scalar fields are both crucial to determine the transition behavior which is reflected in the coupled differential equations Eq.(\ref{inv1})and Eq.(\ref{inv2}). The differential equations may not be directly solved analytically, but in all the cases where Kasner transition happens, we can find numerically that the Kasner transition has two interesting relations
\begin{equation}\label{tr}    \alpha_{1}\beta_{1}+\alpha_{2}\beta_{2}=2, \quad \alpha_{1}\beta_{2}=\alpha_{2}\beta_{1}.
\end{equation}
\begin{figure}[h]
\centering
\includegraphics[width=0.45\textwidth]{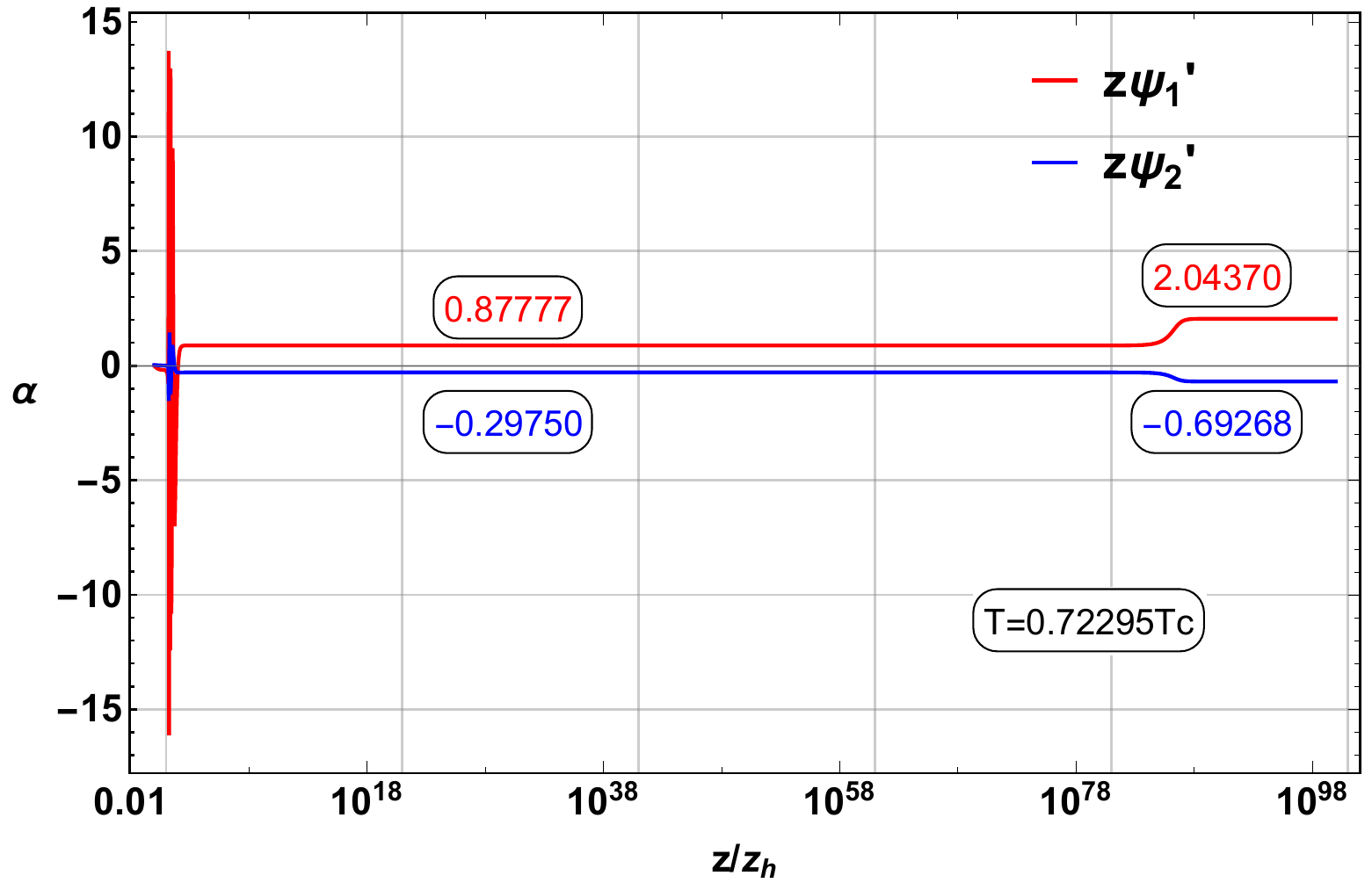}
\caption{Kasner transition behavior for $T=0.72295 Tc$. We find that the exponents before and after the transition satisfy the relation in Eq.(\ref{tr}) }\label{inversion}
\end{figure}
where $\alpha_{1}$ and $\beta_{1}$ are the values of $z\psi'_{1}$ before and after transition, and $\alpha_{2}$ and $\beta_{2}$ are the values of $z\psi_{2}'$ before and after transition. We show one example of this Kasner transition behavior in Fig.\ref{inversion}, and more cases are presented in \ref{more} in Fig.\ref{inversionmore}.  The results are also summarized in the Table \ref{table1}. The relations we find in Eq.(\ref{tr}) are the first Kasner transition rule in the presence of two free parameters. It should be distinguished from other Kasner transition rules in Ref.\cite{An:2022lvo,Cai:2024ltu,Hartnoll:2022rdv} where there is only one free parameter.

\begin{table*}[!h]
\centering
\begin{tabular}{c|c|c|c|c|c|c}   
        \hline
        $T/T_{c}$& $\alpha_{1}$ & $\alpha_{2}$ & $\beta_{1}$ & $\beta_{2}$ &$\alpha_{1}\beta_{1}+\alpha_{2}\beta_{2}$&$\alpha_{1}\beta_{2}-\alpha_{2}\beta_{1}$ \\
        \hline
         0.72295  & 0.87777   &  -0.29750   & 2.04370& -0.69268  &  1.99997 & $5\times 10^{-5}$   \\
       \hline
         0.72292  & 0.71784   & -0.29806  & 2.37641  & -0.98672 & 1.99998 & $2\times 10^{-5} $    \\
        \hline
         0.72275  & -0.09648  & -0.28950    & -2.07215 & -6.21778  & 1.99997 & $2 \times 10^{-5}$   \\
        \hline
         0.72265 & -0.58321 &  -0.27603 & -2.80165 & -1.32601 & 1.99997 & $ 2\times 10^{-6} $  \\
        \hline   
        0.72261 & -0.74294 & -0.27026 & -2.37738 & -0.86482 & 1.99998 & $ 1\times 10^{-6} $  \\
        \hline  
    \end{tabular}
    \caption{\label{table1} The exponents before and after the Kasner transition calculated by numerics. It can be found that the relation in Eq.(\ref{tr}) holds in all cases.}
\end{table*}

It is interesting to check that these two conditions are crucial to make the Kasner transition follow the BKL transition rule \cite{Belinsky:1970ew,Oling:2024vmq,Damour:2002et} which is 
\begin{equation}
    p_{t} \to -\frac{p_{t}}{2p_{t}+1}, \quad p_{x} \to \frac{p_{x}+2p_{t}}{2p_{t}+1}, \quad p_{y} \to \frac{p_{y}+2p_{t}}{2p_{t}+1}.
\end{equation}
We see that $p_{t}$ changes signs during the transition period. Before the transition, $p_{t}$ is negative as a result of $\alpha_{1}^{2}+\alpha_{2}^{2}<2$ while $p_{t}$ is positive after the transition. Therefore, $p_{t}<0$ which physically means that spatial $t$ direction of the geometry expands is an unstable period while $p_{t}>0$ is a stable period. As proved in Ref.\cite{Henneaux:2022ijt}, as long as scalar fields are minimally coupled, there is final Kasner regime no matter how many scalar fields are. So there will be no more Kasner transition after entering into $p_{t}>0$ region. 

\begin{figure*}[h]
    \centering
    \subfigure[]{\includegraphics[width=0.5\textwidth]{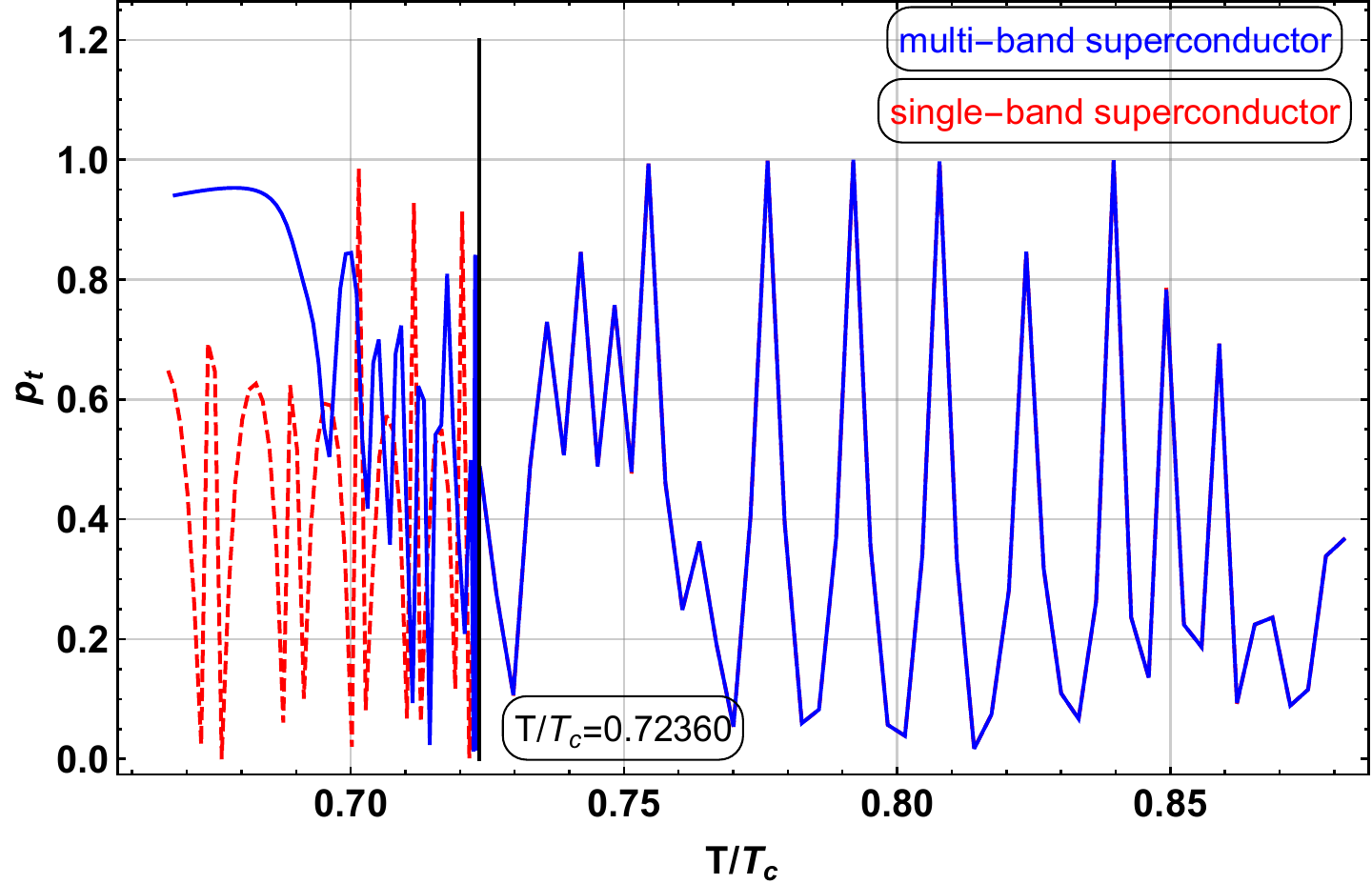}}\hfill
    \subfigure[]{\includegraphics[width=0.49\textwidth]{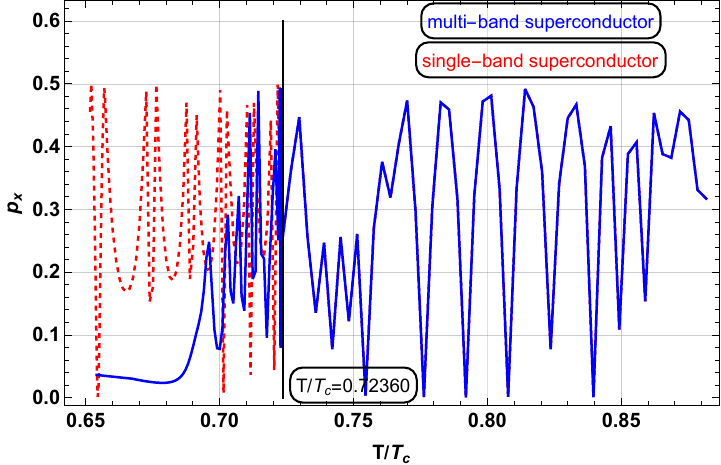}}
    \caption{The Kasner exponents in multi-band superconductor are plotted in blue solid line, from which we see Kasner exponents $p_{t}$ and $p_{x}$ depend sensitively on the temperature. We also plot the Kasner exponents of single band superconductor (with only $\psi_{1}$ field) using red dotted line for comparison. We find that the additional order parameter $\psi_{2}$ will significantly alter the Kasner exponent. }
    \label{pt}
\end{figure*}

Moreover, Kasner exponents $p_{t}$ and $p_{x}$ are also sensitive to the temperature. We plot the relation between the final Kasner exponents and temperature in Fig. \ref{pt}. We find that Kasner exponents $p_{t}$ and $p_{x}$ have irregular behavior in terms of temperature. Furthermore, we also compare the Kasner exponents in multiple order parameters case with the Kasner exponents in single order parameter case. We find that when the second order parameter appears below $T=0.72360T_{c}$, the Kasner exponents will be significantly shifted which means that Kasner exponents near the singularity are sensitive probes of the phases of boundary condensed matter system.

\section{Conclusions and Discussions}\label{cond}
In this work, we investigate the interior structure of holographic multi-band superconductor.  For the region near the horizon, there still exists ER collapse behavior after which Josephson oscillation occurs. When lowering the temperature below which second scalar condensate appears, there will be oscillation behavior of two scalar fields. For the region near the singularity, the additional scalar field has contributions to the Kasner exponent which makes the Kasner exponent different from the single scalar case. In this multi-band case, the Kasner exponents are determined by two independent parameters $\alpha_{1}$ and $\alpha_{2}$ which is more complex than the single-band case. This feature affects the Kasner transition behavior significantly, which makes the Kasner transition present distinct relations compared to the single band case. 
The transition rule we found in Eq.(\ref{tr}) is the first kind of Kasner transition in the presence of two independent parameters which will be crucial for characterizing the general interior structure of hairy black holes. 
Furthermore, these results certainly illuminate that the black hole interior plays crucial role in describing the boundary condensed matter system. 

In this work, we only focus on the s-wave superconductor. However, p-wave superconductor can have more intricate interior structure compared to s-wave case \cite{Cai:2021obq}. So it will be interesting to investigate the interior of multi-band model with coexistence of s-wave and p-wave order parameters \cite{Nie:2014qma,Nie:2013sda} which may present more novel phenomena. 

Moreover, 
it is interesting to ask if the near singularity structure we found in Fig.\ref{pt} can be directly reflected in some physical quantities of boundary systems. Based on the proposal raised in \cite{Caceres:2022smh}, we can claim that the structure we found can at least be closely related to the renormalization group theory of boundary system.

For AdS black hole, the radial coordinate $z$ is commonly treated as an energy scale which parameterized the renormalization group flow from the UV CFT on the boundary to the IR CFT on the horizon. It was shown in \cite{Caceres:2022smh} that by analytically continuation, the holographic renormalization group flow can also be generalized to describe the black hole interior region, which is called trans-IR flow. Along the flow from UV to IR and then to trans-IR region, a monotonic function can be defined which is called thermal a-function \cite{Caceres:2022smh}. The thermal a-function measures the effective degrees of freedom of field theory at each energy scale along the RG flow whose expression in $d+1$ dimensions is 
\begin{equation}
    a_{T}(u)=\frac{\pi^{d/2}}{\Gamma(\frac{d}{2})\kappa^{d-1}}(\frac{\sqrt{f(u)}e^{-\chi/2}}{A'(u)})^{d-1},
\end{equation}
where $\kappa$ is the Planck length and 
\begin{equation}
    \frac{du}{dz}=-\frac{1}{z\sqrt{f}}, \quad e^{2A(u)}=\frac{1}{z^{2}}.
\end{equation}
Note that the thermal a-function reduces to central charge of boundary CFT for $z \to 0$. Moreover, the monotonicity of thermal a-function can be proved by using null energy condition. 

In four dimensional case, the thermal a-function simplifies to be \footnote{Note that the monotonically decreasing behavior of thermal a-function can also been seen from the monotonically increasing behavior of $\chi$ field.}
\begin{equation}
    a_{T}(z)=\frac{2\pi}{\kappa^{2}}e^{-\chi}.
\end{equation}
Near the singularity, we found that the thermal a-function becomes 
\begin{equation}
    a_{T}(z)|_{\mathrm{kasner}}=\frac{2\pi}{\kappa^{2}}e^{-(\alpha_{1}^{2}+\alpha_{2}^{2})\ln z}=\frac{2\pi}{\kappa^{2}} z^{-\frac{6p_{t}+2}{1-p_{t}}},
\end{equation}
where we use Eq.(\ref{chik}). Thus we see that the thermal a-function in Kasner region is solely determined by the Kasner exponent $p_{t}$ and different behavior of Kasner exponent will lead to different behavior of thermal a-function (or different behavior of effective degrees of freedom). So the thermal a-function is a good physical quantity to reflect the different near singularity structure between multi-band superconductor model and single band superconductor model. Besides thermal a-function, there are also proposals using correlation functions \cite{Frenkel:2020ysx,Caceres:2023zft,Grinberg:2020fdj,David:2022nfn} and complexity \cite{Jorstad:2023kmq} to characterize the near singularity structure of black hole interior, we leave the analysis of these variables to future investigations.
 
\section*{Acknowledgements}
We are grateful for the useful discussions with our group members. This work is supported by the National Natural Science Foundation of China (NSFC) under Grant Nos.12405066, 12175105 and 11965013. YSA is also supported by the Natural Science Foundation of Jiangsu Province under Grant No. BK20241376 and Fundamental Research Funds for the Central Universities. 


\appendix

\section{More Kasner transition diagram:}\label{more}
\begin{figure*}[h]
    \centering
    \subfigure[]{\includegraphics[width=0.5\textwidth]{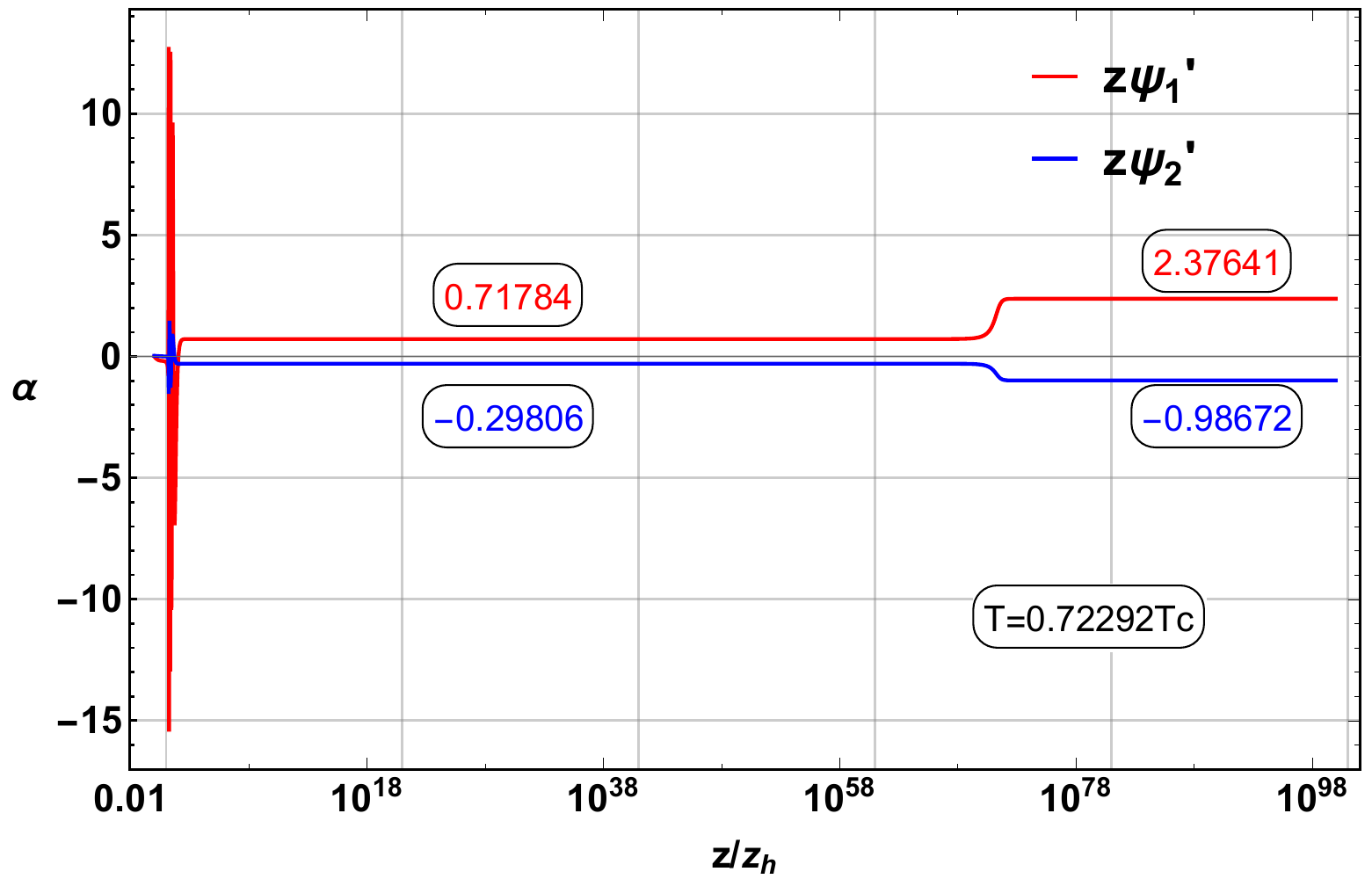}}\hfill
    \subfigure[]{\includegraphics[width=0.5\textwidth]{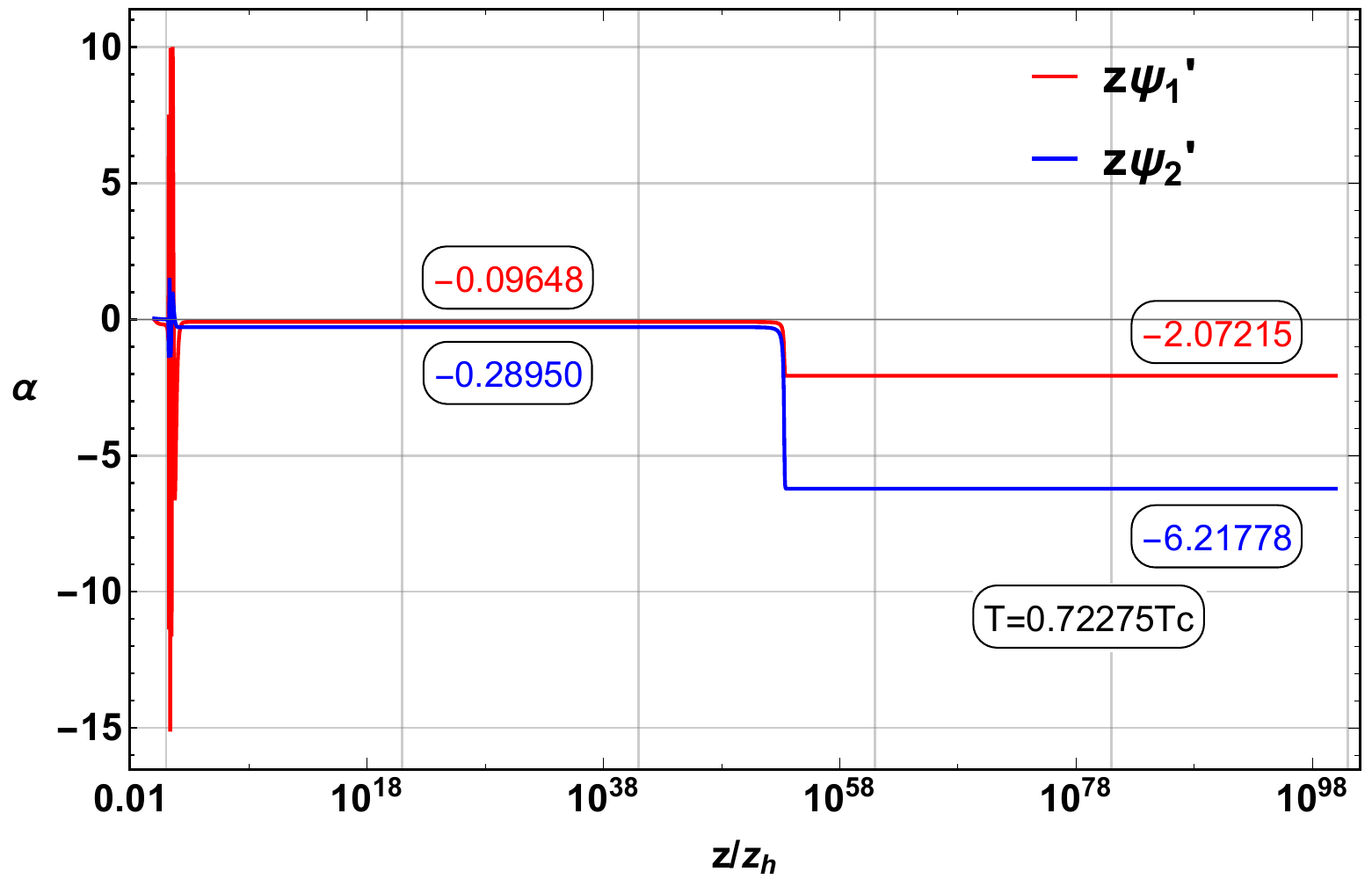}}
     \subfigure[]{\includegraphics[width=0.49\textwidth]{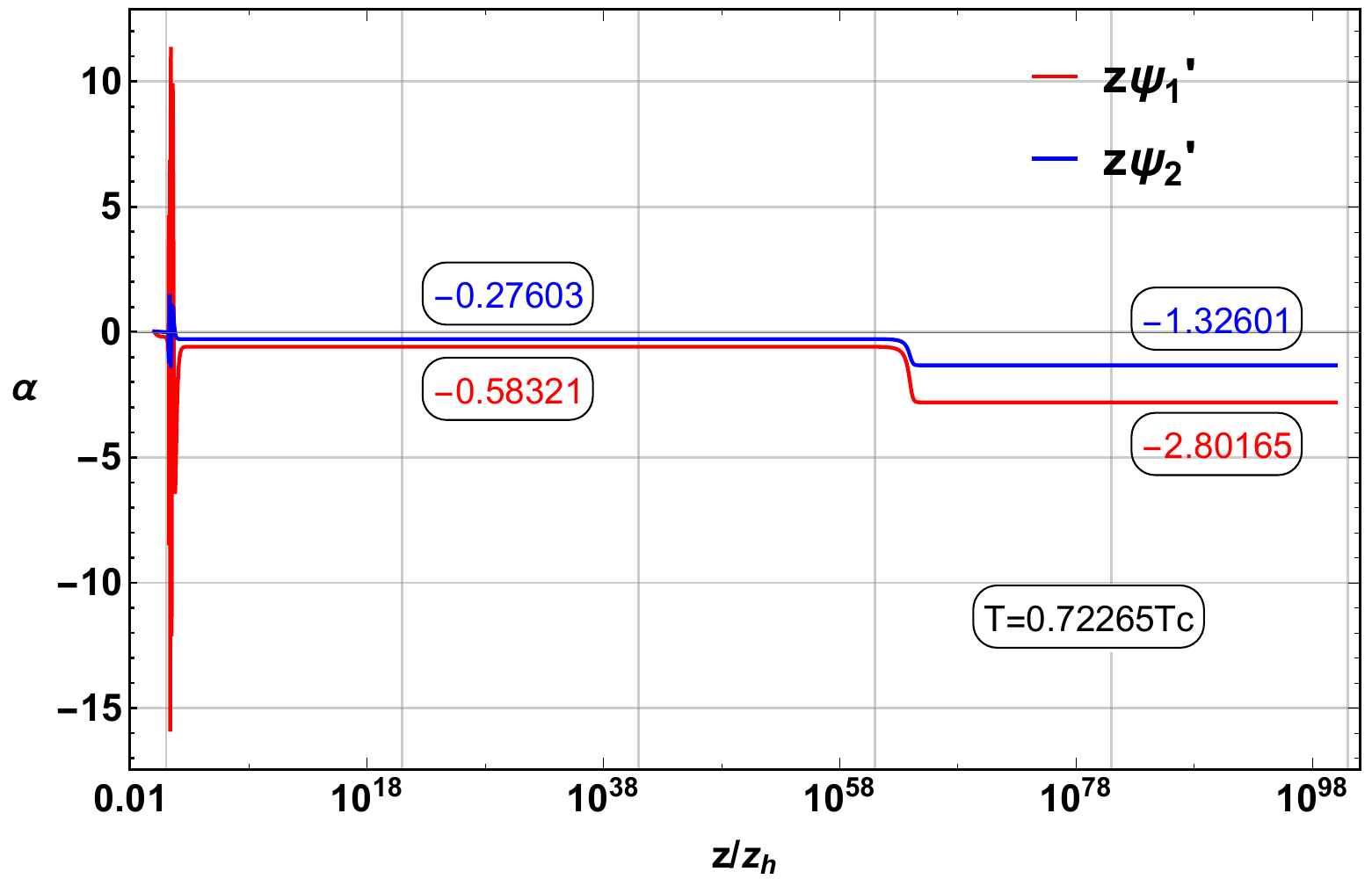}}
      \subfigure[]{\includegraphics[width=0.49\textwidth]{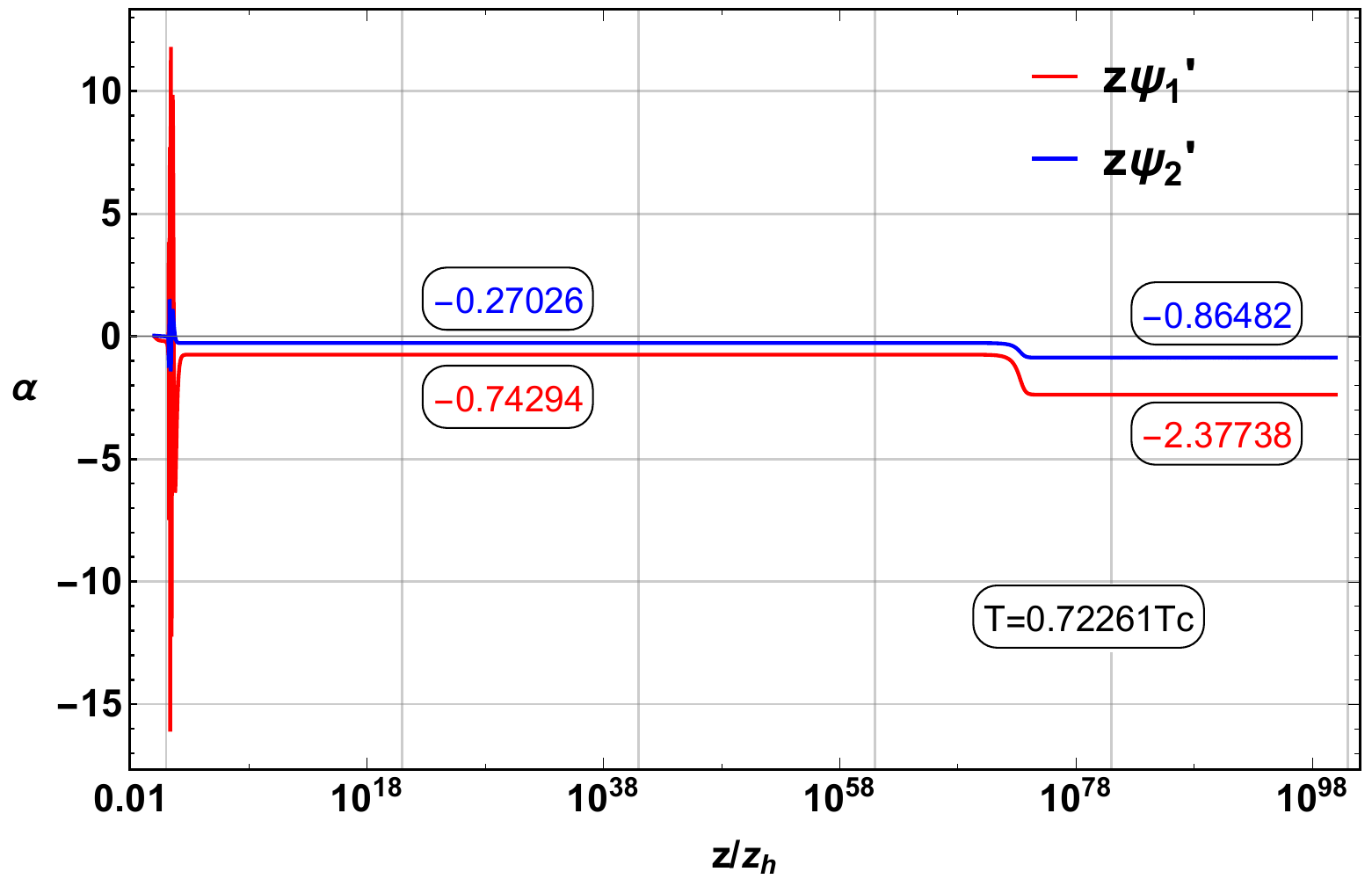}}
    \caption{Kasner inversion behavior for  $T=0.72292 Tc$, $T=0.72275 Tc$, $T=0.72265 Tc$ and $T=0.72261 T_{c}$, all of these example satisfy the two relations in Eq.(\ref{tr}). }
    \label{inversionmore}
\end{figure*}

In this appendix, we show more examples where the Kasner epochs have transition behavior in Fig.\ref{inversionmore}, these transition behaviors all satisfy the relation we found in Eq.(\ref{tr}) thus obeying BKL transition relation. 

\bibliographystyle{elsarticle-num} 
\biboptions{sort&compress}
\bibliography{msc}

\end{document}